\documentclass[aps,twocolumn,preprintnumbers,nofootinbib,floatfix]{revtex4-1}
\usepackage{dcolumn}
\usepackage{bm}
\usepackage{epsfig}
\usepackage{graphicx}
\usepackage{amsmath}
\usepackage{amssymb}
\usepackage{mathrsfs}
\usepackage{color}
\usepackage[usenames,dvipsnames]{xcolor}
\usepackage{hyperref}
\usepackage[caption=false]{subfig}

\def\be{\begin{equation}}
\def\ee{\end{equation}}
\def\bea{\begin{eqnarray}}
\def\eea{\end{eqnarray}}

\newcommand{\comments}[1]{}

\begin{document}

\title{\vspace*{-0.5cm}Hints of Entanglement Suppression in Hyperon-Nucleon Scattering}

\author{\vspace{-0.2cm}
{Qiaofeng Liu$^{1}$ and Ian Low$^{1,2}$}
 }
\affiliation{\vspace*{0.1cm}
$^1$\mbox{\small Department of Physics and Astronomy, Northwestern University, Evanston, IL 60208, USA} \\
$^2$\mbox{\small High Energy Physics Division, Argonne National Laboratory, Argonne, IL 60439, USA}\\
\vspace*{-0.5cm}
}

\begin{abstract}
Hyperon ($Y=\Sigma,\Lambda$)-nucleon ($N=n,p$)  interactions are crucial for understanding the existence of  neutron stars heavier than two solar masses. Amid renewed experimental efforts, we study  $YN$ scatterings from the perspective of quantum information, focusing on whether  spin entanglement is suppressed in the s-wave channel, which is observed in $np$ scattering and leads to enhanced global symmetries. Using global fits of phase shifts from experimental  data, we find hints of entanglement suppression among the eight flavor channels in the strangeness $S=-1$ sector, similar to the $np$ case. One exception is the $\Sigma^+p$ channel, where conflicting global fits lead to inconclusive outcome. We  then propose ``quantum'' observables  in  $\Sigma^+p$ scattering to help resolve the differing global fits. 
\end{abstract}

\maketitle

\section{Introduction} 
Understanding the dynamics of baryons is essential for obtaining a comprehensive picture of strong interactions, with broad implications in particle, nuclear, and astrophysics.  While nucleon-nucleon  interactions have been measured and constrained very precisely based on more than 6700 data points \cite{Stoks:1993tb,Wiringa:1994wb,gross2008covariant,machleidt2001high,perez2013partial}, interactions involving baryons carrying strangeness quantum number -- the hyperon  -- are much less understood, due to experimental difficulties associated with  the short lifetime of  hyperon and the challenges of preparing a stable hyperon beam. Very little experimental progress was made on $YN$ scattering since the 1970s, until recently when the E40 collaboration at J-PARC, the CLAS collaboration at CEBAF and the BESIII collaboration at BEPCII announced new measurements \cite{J-PARCE40:2021qxa,J-PARCE40:2021bgw,J-PARCE40:2022nvq,CLAS:2021gur, BESIII:2009fln,BESIII:2023clq}. Furthermore, there are also relevant measurements on $YN$ interactions at the Large Hadron Collider from the ALICE collaboration recently \cite{ALICE:2019buq,ALICE:2021njx}.

These new experimental efforts are driven, in part, by the  ``hyperon puzzle'' in understanding the existence of neutron stars with a mass larger than two solar-mass \cite{Chatterjee:2015pua,Tolos:2020aln}. Neutron stars are  compact  objects supported against gravitational collapse by the Fermi degeneracy pressure of the neutron. However, in a dense environment  the hyperon becomes stable because of the limited decay phase space and the Fermi pressure is  reduced by the conversion of neutrons into hyperons due to the large chemical potential  inside the neutron star, thereby softening the equation-of-state (EOS). Such a  feature turns out to be incompatible with the observation of a neutron star heavier than two solar-mass \cite{Demorest:2010bx,Antoniadis:2013pzd}  as well as the stiff EOS measured by  LIGO-Virgo  \cite{LIGOScientific:2018hze,LIGOScientific:2018cki}.

There has been continuous theoretical effort in describing $NN$, $YN$ and $YY$ interactions over the years. Some  popular approaches include meson-exchange potential models (the Nijmegen \cite{Rijken:1998yy,Stoks:1999bz,Rijken:2010zzb} and the J\"ulich \cite{Holzenkamp:1989tq,Haidenbauer:2005zh} potentials), the chiral effective field theory ($\chi$EFT) approach \cite{Polinder:2006zh,Haidenbauer:2013oca,Haidenbauer:2019boi}, and lattice QCD simulations (see the HALQCD \cite{Ishii:2006ec,Nemura:2017vjc} and the NPLQCD \cite{Beane:2006gf,NPLQCD:2020lxg} collaborations). The underlying organizing principle is the $SU(3)_f$ flavor symmetry among $(u,d,s)$ quarks, which  is broken by quark masses. The spin-$1/2$ baryons, $(n, p, \Sigma^\pm, \Sigma^0, \Lambda, \Xi^0, \Xi^-)$,  fill out an octet under  $SU(3)_f$. A major  challenge is  to incorporate $SU(3)_f$ breaking effects in parameterizing $YN$ and $YY$ interactions  amid constraints from the $NN$ sector, and  then extract useful information by fitting to the scarce data. Not surprisingly, ambiguities and inconsistencies among different approaches remain. Novel experimental techniques \cite{Dai:2022wpg} and new theoretical insights are urgently needed.

Parallel to recent experimental advances, a new theoretical perspective  was proposed in Ref.~\cite{Beane:2018oxh},  which utilizes information-theoretic tools to analyze emergent  symmetries in low-energy QCD. It was discovered that the S-matrix for $np$ scattering below the pion threshold tends to suppress the spin-entanglement throughout the scattering, and such a suppression correlates with the emergence of Wigner's $SU(4)_{sm}$ spin-flavor symmetry \cite{Mehen:1999qs} and the Schr\"odinger invariance \cite{Mehen:1999nd} in the $NN$ sector. Assuming $SU(3)_f$ symmetry, Ref.~\cite{Beane:2018oxh} also studied  the scattering between spin-1/2 baryons and found that entanglement suppression could give rise to an emergent $SU(16)$ spin-flavor symmetry. 

Building on the new insights, Ref.~\cite{Low:2021ufv} analyzed the $s$-wave scattering of non-relativistic fermions from an information-theoretic setting and showed that the spin-flavor symmetry is associated with the S-matrix being interpreted as an Identity quantum logic gate and the Schr\"odinger symmetry is related to the SWAP gate. 
In Ref.~\cite{Liu:2022grf} pathways to other emergents symmetries, such as $SU(6)$, $SO(8)$ and $SU(8)$, in baryon-baryon scatterings were pointed out. While the aforementioned physical systems are non-relativistic,  recently the correlation between entanglement  and  symmetries was extended to a fully relativistic quantum system involving two-Higgs-doublet models and electroweak symmetry breaking, where a standard-model-like Higgs boson arises as a consequence of entanglement suppression \cite{Carena:2023vjc}.

Inspired by the advances in both experiment and theory, in this work we pursue a study on  $YN$ scattering from the viewpoint of quantum information. Our goal is two-fold: constrain the information-theoretic property of low-energy $YN$ scattering using global fits of current  data and propose  quantum information-sensitive observables which may  improve our understanding. Using the global fits is complementary to lattice QCD simulations, which  assume  $SU(3)_f$ symmetry and sometimes adopt an unrealistic pion mass due to  limited computing powers.

\section{Kinematics}

In the experimental setup, a beam of hyperon particles hit a stationary target, the nucleon. We label the incident particle mass as $m_1$ and the stationary particle mass $m_2$. Masses of outgoing particles are represented by $m_1'$ and $m_2'$. Experimental observables are usually measured as a function of the kinematic energy $T_{lab}$ and the magnitude of the 3-momentum ${P}_{lab}$ of the incident particle in the laboratory (Lab) frame. We center on 2-to-2 scattering and therefore focus on the kinematic regime below the pion production threshold, 
\begin{equation}
	N_1 + N_2\to N_1' + N_2' + \pi \ .
\end{equation}
It is the easiest to calculate the pion threshold in the centre-of-mass (CM) frame which  in terms of $p_{CM}$, the magnitude of the 3-momentum of the incident particle, is 
\begin{equation}
	{p}_{CM}^2 = \frac{\left[m_1'm_2'+(m_1'+m_2'+m_\pi/2)m_\pi\right]^2-m_1^2m_2^2}{(m_\pi+2m_1'+2m_2')m_\pi+(m_1+m_2)^2}\ ,
\end{equation}
where $m_\pi$ is the pion mass. In Table.~\ref{table:pion-threshold} we list the representative pion production thresholds, in both the CM and the Lab frame, for $YN$ scattering. The kinematic threshold in $p_{CM}$ spans  between $(380,390)$ MeV/c.
\begin{table}[t]
	$\begin{array}{c|c|c}
		\hline
		\text{Pion production process}&\text{$p_{CM}$ } (\text{MeV/c}) & \text{$p_{lab}$ } (\text{MeV/c})   \\
		\hline
		\Lambda n \to \Lambda p \pi^-& 382.8&893.9\\
		\Sigma^+ p \to \Sigma^+ n \pi^+& 390.3&943.4\\
		\hline
	\end{array}$   
	\caption{\em Representative pion production thresholds for  $YN$.}\label{table:pion-threshold}
\end{table}

One interesting feature of $YN$ scattering is the outgoing particles could have different flavors from the incoming particles, the flavor non-diagonal channels, in contrast with the $NN$ scattering which is always flavor diagonal. In the limit of exact $SU(3)_f$, all spin-1/2 octet baryons are degenerate in mass and, since the strong interaction preserves the electric charge $Q$ and the strangeness $S$, flavor channels within the same $(Q,S)$ sector scatter only among themselves. For $S=-1$ sector the  flavor channels are classified according the electric charge $Q$ in Table \ref{table:flavor-subspace}. Notice that $\Sigma^- n$ and $\Sigma^+ p$ are unique in their respective $(Q, S)$ sector, whose scatterings are always elastic and flavor diagonal.

\begin{table}[t]
	$\begin{array}{l|c|c|c|c}
	\hline
	Q & -1 & 0  & 1 & 2 \\
	\hline
	\text{Flavor} &   \Sigma^- n&  \Lambda n, \Sigma^0 n, \Sigma^- p& \Lambda p, \Sigma^0 p, \Sigma^+n & \Sigma^+ p\\
	\hline
	\text{Total} & 2137 & \Lambda n : 2055  & \Lambda p: 2054  & 2128   \\
	\text{Mass (MeV)}	& & \Sigma^0 n : 2132  & \Sigma^+ n : 2129 &      \\
		& & \Sigma^- p : 2136 & \Sigma^0 p : 2131 &    \\
		\hline
		\end{array}$   
	\caption{\em Eight flavor channel and the total mass in the strangeness $S=-1$ sector, as labeled by the total charge $Q$. 
	}\label{table:flavor-subspace}
\end{table}

In reality, $SU(3)_f$  is broken  and the baryons are not  degenerate in mass. In Table \ref{table:flavor-subspace} we also show the total  mass of each $YN$ channel, from which we see $\Lambda p$ and $\Lambda n$ are the lightest flavor channel in the $Q=0$ and $Q=1$ sectors respectively. This implies  $\Lambda p$ and $\Lambda n$ would scatter elastically and flavor-diagonally until the kinematic thresholds for the next lightest flavor channels open up, which in $(p_{CM}, P_{lab})$  are $(279,633)$ MeV/c  for $\Lambda p$ and $(283,633)$ MeV/c for $\Lambda n$, respectively.

\section{The S-matrix}

It is well-known that non-relativistic scatterings of spin-1/2 fermions in the low-energy are dominated by the $s$-wave channel, which contains the spin-singlet  $^1S_0$  and the spin-triplet $^3S_1$ \cite{Low:2021ufv},
\begin{equation}
\label{eq:smatpspt}
S= e^{2i \delta_0} P_{s} + e^{2i \delta_1} P_{t} \ ,  
\end{equation}
where $P_s$ and $P_t$ are  the spin-projectors into the $^1S_0$ singlet and the $^3S_1$ triplet channels, respectively, 
\begin{equation}
P_{s}=\frac14\left(1-\bm{\sigma}\cdot\bm{\sigma}\right) \ , \qquad
 P_{s} =\frac14\left(3+\bm{\sigma}\cdot\bm{\sigma}\right), 
\end{equation} 
and $\bm{\sigma}\cdot\bm{\sigma} = \sum_a \sigma^a\otimes \sigma^a$. Moreover, $\delta_0$ and $\delta_1$ are the scattering phase shifts which can be fitted from  data.

The information-theoretic property of the S-matrix becomes transparent when we introduce the SWAP operator, ${\rm SWAP}= (1+\bm{\sigma}\cdot\bm{\sigma})/2$, which interchanges the spin of the two  particles, and rewrite the S-matrix   as  \cite{Low:2021ufv},
\begin{equation}
S= \frac12\left(e^{2i \delta_0}+e^{2i \delta_1}\right) 1 +\frac12 \left(e^{2i \delta_0}-e^{2i \delta_1}\right) {\rm SWAP} \ .
\end{equation}
It then becomes clear that the S-matrix in the spin subspace is an Identity operator in the spin-space if $\delta_0=\delta_1$ while it is the SWAP operator if $|\delta_0-\delta_1|=\pm \pi/2$. It turns out that 1 and SWAP are the only two operators which   suppress entanglement \cite{Low:2021ufv}, which can be seen from the observation that, starting from an arbitrary spin-wave function $|\psi\rangle$ that is unentangled, $|\psi\rangle=|\psi_1\rangle\otimes |\psi_2\rangle$, the Identity and the SWAP operators both produce an outgoing state that is also unentangled. Note that  ${\rm SWAP}|\psi_1\rangle\otimes |\psi_2\rangle=|\psi_2\rangle\otimes |\psi_1\rangle$. It is much less obvious  that these are the only two two-qubit logic gates which suppress entanglement \cite{Low:2021ufv}. For a unitary operator, it is possible to quantify the ability of an operator to generate entanglement by defining the entanglement power (EP) \cite{BallardWu2011,PhysRevA.70.052313}, which for the S-matrix  in Eq.~(\ref{eq:smatpspt}) is \cite{Beane:2018oxh},
\begin{equation}
\label{eq:epphase}
E(S) = \frac16 \sin^2(2\delta_0-2\delta_1) \ .
\end{equation}
It vanishes when $\delta_0=\delta_1$ (the Identity) or $|\delta_0-\delta_1|=\pm \pi/2$ (the SWAP). 

For a non-unitary operator, however, the EP is not well-defined. This is the case when we consider inelastic $YN$ scattering in the $Q=0$ and $Q=1$ sectors, where the S-matrix  is three-dimensional in the flavor subspace, as can be seen in Table \ref{table:flavor-subspace}. Therefore,  the flavor-diagonal entry of the S-matrix by itself is not a unitary   operator and it is customary to introduce the inelasticity parameter, which can also be extracted from data, for each flavor-diagonal channel \cite{Sprung:1985zz},
\begin{equation}
\label{eq:smatnon}
[S]_{jj} = \cos(2\alpha^j_0) e^{2i \delta^j_0} P_{s} +\cos(2\alpha^j_1)  e^{2i \delta^j_1} P_{t} \ ,
\end{equation}
where $[S]_{jj}$ is the diagonal entry of S-matrix describing the scattering $j\to j$ in the flavor subspace. Below the inelastic threshold the inelasticity parameters vanish, which is the case  for $\Lambda p$ and $\Lambda n$ channels. Although the EP is not well-defined for $[S]_{jj}$  in Eq.~(\ref{eq:smatnon}), one can still rewrite $[S]_{jj}$ as
\begin{align}
\label{eq:smatnon1}
[S]_{jj} &= c_1 \ 1 + c_{\rm S}\ {\rm SWAP}  \ , \\
\label{eq:smatnon2}
c_1 &= \frac12\left[\cos(2\alpha^j_0) e^{2i \delta^j_0}+\cos(2\alpha^j_1) e^{2i \delta^j_1}\right] \ ,\\
\label{eq:smatnon3}
c_{\rm S}&= \frac12 \left[\cos(2\alpha^j_0)e^{2i \delta^j_0}-\cos(2\alpha^j_1) e^{2i \delta^j_1}\right]  .
\end{align}
Then $[S]_{jj}$ suppresses spin entanglement when the inelasticity parameters and the phase shifts 
conspire in such a way that $[S]_{jj}\sim 1$ or $[S]_{jj}\sim$ SWAP. 

\section{Global Fits}

There are two types of global fits in the literature. One type provides the global fit for the  momentum dependence of the phase shifts $\delta_0(p)$ and $\delta_1(p)$ in the $^1S_0$ and $^3S_1$ channels, from which we  calculate the EP using Eq.~(\ref{eq:epphase}). 
In particular, we focus on two different fits developed for different purposes: the Nijmegen soft-core model (NSC97) \cite{Rijken:1998yy,Stoks:1999bz,nnonline} and the Nijmegen extended-soft-core model (ESC16) \cite{Nagels:2014qqa,Nagels:2015lfa}. 

\begin{figure}[t]
	\centering
	\includegraphics[width=.23\textwidth]{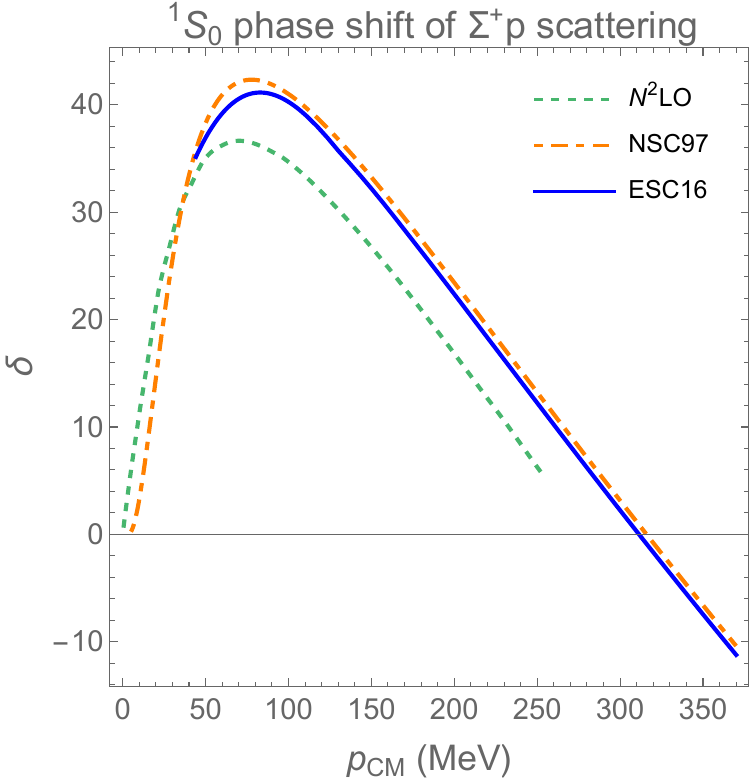}
		\includegraphics[width=.23\textwidth]{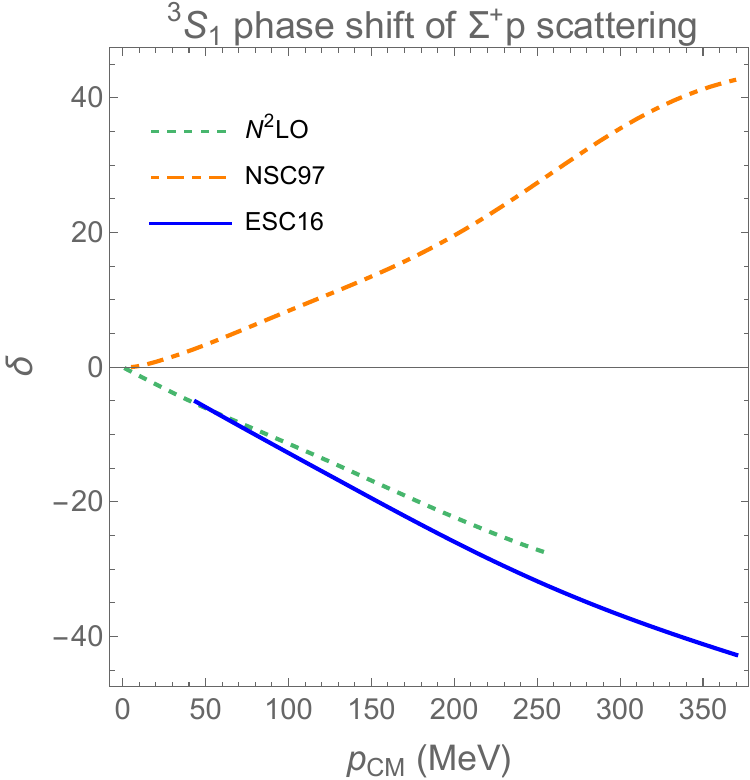}
	\caption{\em Comparisons of phase shifts in the $\Sigma^+ p$ channel from potential models and N$^2$LO $\chi$EFT ($\Lambda=$500 MeV).}
	\label{fig:phasecompare}
\end{figure}

The main difference between these two models lies in the $\Sigma^+p$ channel, the $(Q, S)=(2,-1)$ and isospin $I=3/2$ sector, where the NSC97 model predicts an attractive force  in the $^3S_1$ channel while the ESC16 model predicts a repulsive force. Historically two-body $\Sigma^+p$ interactions are fitted from binding energies of hypernuclei, which seem to prefer a strong repulsive interaction  \cite{Rijken:2010zzb,Hashimoto:2006aw}. However, extracting the two-body interaction from hypernuclear data requires knowledge of the three-body $YNN$ force, which  plays an important role in understanding the hyperon puzzle inside the  neutron star \cite{Yamamoto:2014jga} and is not precisely determined. In the end, it is highly desirable to be able to independently determine the two-body $\Sigma^+p$ interactions directly from scattering data. In this regard, the latest phase shift analysis from measurements at J-PARC does not seem to be able to resolve the difference between ESC16 and NSC97 \cite{J-PARCE40:2022nvq}. Consequently we utilize both models in the present work. We will see that these two models make distinctive predictions on the entanglement property of $\Sigma^+p$ scattering.

The second type of fits makes use of $\chi$EFT, which is pioneered by Weinberg \cite{Weinberg:1990rz,Weinberg:1991um} and has been very successful in describing $NN$ interactions. Applications of $\chi$EFT to $YN$ interactions have progressed steadily over the years \cite{Polinder:2006zh,Haidenbauer:2013oca,Haidenbauer:2019boi} and the state-of-the-art calculation now stands at next-to-next-to-leading order (N$^2$LO) \cite{Haidenbauer:2023qhf}, which we use.

There is one feature of $\chi$EFT which is distinctly different from the global fits utilizing the potential models such as NSC97 and ESC16. $\chi$EFT is an expansion of the potential in small momenta and pion masses, augmented with an appropriate power counting rule. As such there is an inherent cutoff of $\chi$EFT, which is usually taken to be $\Lambda \sim 500$ MeV. This implies that,  as the $p_{CM}$ gets close to $\Lambda$, higher order effects become important and $\chi$EFT starts breaking down. 
In Fig.~\ref{fig:phasecompare} we show a comparison of the phase shifts in the $\Sigma^+p$ channel from the potential models and from $\chi$EFT. 
We see that phase shifts from ESC16 and N$^2$LO are consistent up to $p_{CM}\sim 250$ MeV.  
Moreover, the $^3S_1$ phase shifts in ESC16 and N$^2$LO $\chi$EFT have a negative sign (repulsive force) while in NSC97 it has a positive sign (attractive force).  In the isospin-related channel of $\Sigma^- n$, lattice simulations seem to also indicated a negative $^3S_1$ phase shift \cite{Beane:2012ey}.

\section{Results}

\begin{figure}[t]
	\centering
	\includegraphics[width=.23\textwidth]{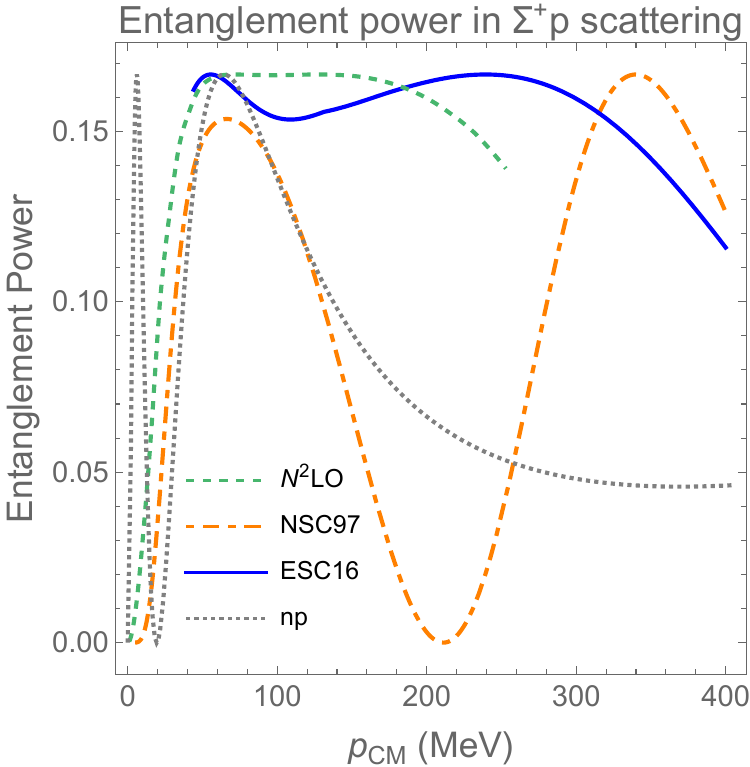}		\includegraphics[width=.23\textwidth]{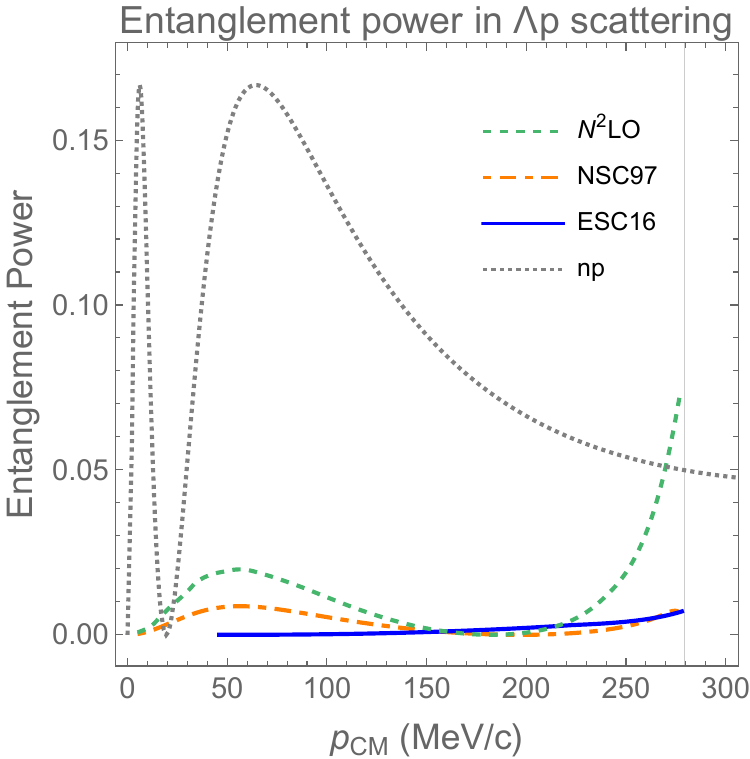}
	\caption{\em EP in $\Sigma^+ p$ and $\Lambda p$ channels, which are related  to $\Sigma^- n$ and $\Lambda n$ channels, respectively, by isospin invariance. The EP for $np$ scattering is included for comparison.}
	\label{fig:class1ep}
\end{figure}

\begin{figure}[t]
	\centering
	\includegraphics[width=.23\textwidth]{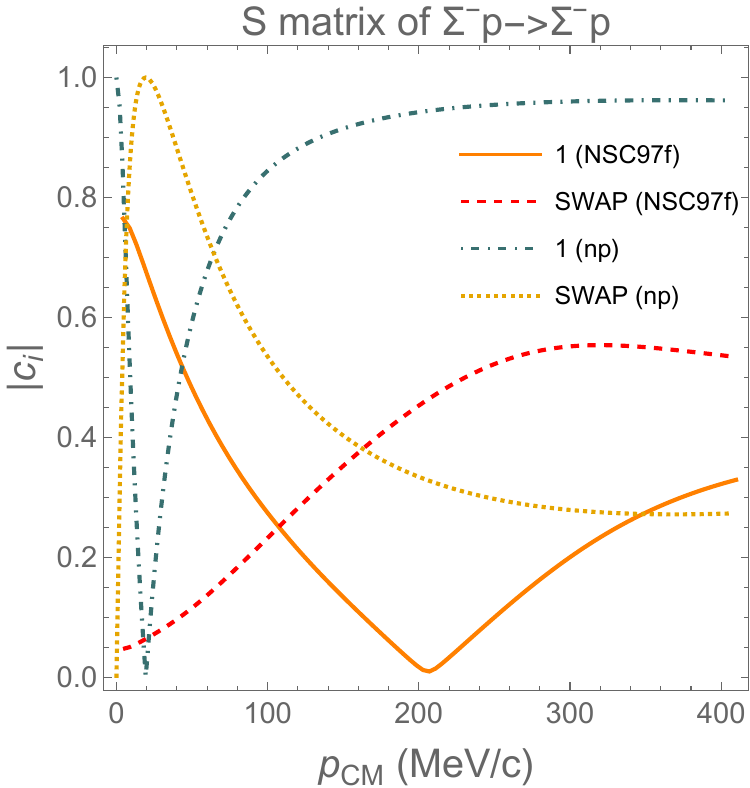}
	\includegraphics[width=.23\textwidth]{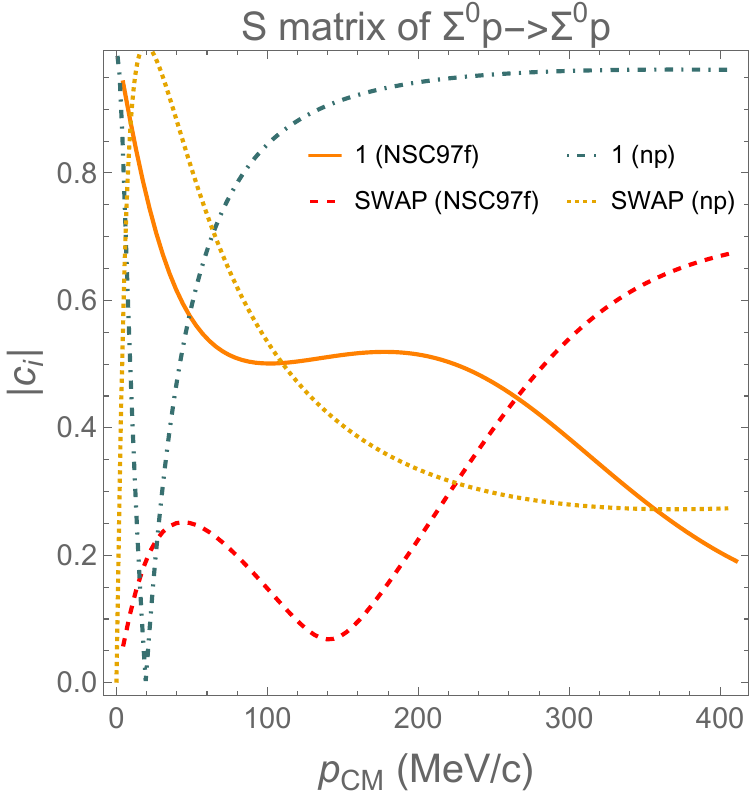}
	\caption{\em S matrices in $\Sigma^- p$ and $\Sigma^0 p$ channels, which are related to $\Sigma^+ n$ and $\Sigma^0 n$, respectively, by isospin invariance. The  S-matrix for $np$ scattering is also shown.}
	\label{fig:class2Smat}
\end{figure}

Here we present the information-theoretic properties of two-body $YN$ interactions. There are eight flavor channels, as shown in Table \ref{table:flavor-subspace}. Among them we will only show the results for $\{\Sigma^+ p, \Sigma^0 p, \Sigma^- p, \Lambda p\}$, since isospin invariance relates them to the remaining four channels, $\{\Sigma^-n, \Sigma^0 n, \Sigma^- n, \Lambda n\}$, and the results look very similar. Notice that, in the ultra-low momentum region where non-perturbative structures such as poles and resonances dominate,  the S-matrix could produce highly entangled states. Our interest lies in the regime above this infrared region,  $p_{CM}\agt 100$ MeV, where nucleons and baryons can be considered as fundamental degrees of freedom.

There are two classes of results. The first  involves elastic scatterings in the flavor-diagonal channels. They are $\Lambda n$ and $\Lambda p$ below the inelastic threshold at $p_{CM}\sim 280$ MeV, as well as $\Sigma^+ p$ and $\Sigma^- n$ below the pion production threshold at $p_{CM}\sim 390$ MeV. (See Tables \ref{table:pion-threshold} and \ref{table:flavor-subspace}.) In Fig.~\ref{fig:class1ep} we present the EP computed from Eq.~(\ref{eq:epphase}) for $\Sigma^+ p$ and $\Lambda p$ channels, and include  the case  of $np$ scattering for comparison. We employ three global fits: NSC97, ESC16 and $\chi$EFT. In particular, NSC97 fits contain several versions \cite{Rijken:1998yy,Stoks:1999bz}, among which we choose the NSC97f as a representative, although the conclusion does not depend on this choice.  We see in Fig.~\ref{fig:class1ep} that the EP is highly suppressed in the region of  $p_{CM}\agt 100$ MeV in the NSC97 fit for $\Sigma^+ p$ and in all three fits for $\Lambda p$.  ESC16 and $\chi$EFT do not exhibit  entanglement suppression in the $\Sigma^+ p$ channel due to the negative $^3S_1$ phase shift, as shown in Fig.~\ref{fig:phasecompare}. In the $\Sigma^-n$ and $\Lambda n$ channels, only the NSC97 fit is  available and the EP is suppressed in both channels, similar to their isospin partners.

The remaining channels, $\{\Sigma^- p, \Sigma^0 p, \Sigma^+ n,\Sigma^0 n\}$,  all scatter inelastically and we need to include the inelasticity parameters. In these cases we plot $|c_1|$ and $|c_{\rm S}|$ in Eqs.~(\ref{eq:smatnon2}) and (\ref{eq:smatnon3}), which show the relative component of the S-matrix in 1 and SWAP. When the S-matrix is predominantly the Identity gate, or the SWAP gate, it suppresses entanglement. In Fig.~\ref{fig:class2Smat} we present the results for $\Sigma^- p$  and $\Sigma^0 p$ channels. The results  for their isospin partners $\Sigma^+ n$ and $\Sigma^0 n$ look very similar. We again show the $np$ case as a benchmark. Only the  NSC97 fit is shown here, as the $\chi$EFT only provides fits for the scattering length, without the effective range \cite{Haidenbauer:2023qhf}. We see that in $\Sigma^- p$  channel the S-matrix is dominated by the SWAP gate at $p_{CM}\sim 200$ MeV, while in $\Sigma^0 p$  channel it is dominated by the Identity gate at $p_{CM}\sim 150$ MeV. Since   the SWAP gate is associated with the Schr\"odinger symmetry \cite{Low:2021ufv},  it would be interesting to further investigate   the appearance of SWAP gate in these channels.

It is worth emphasizing that, even if one assumes $SU(3)_f$ symmetry,  these hints of entanglement suppression in $YN$ interactions do not follow from the observed entanglement suppression in $np$ scattering. In this limit, there are six $SU(3)_f$ invariant phase shifts in the scattering of octet baryons and the phases entering $np$ scattering are different from those appearing in $YN$ scattering \cite{Liu:2022grf}. 

\begin{figure}[t!]
	\centering
	\includegraphics[width=.23\textwidth]{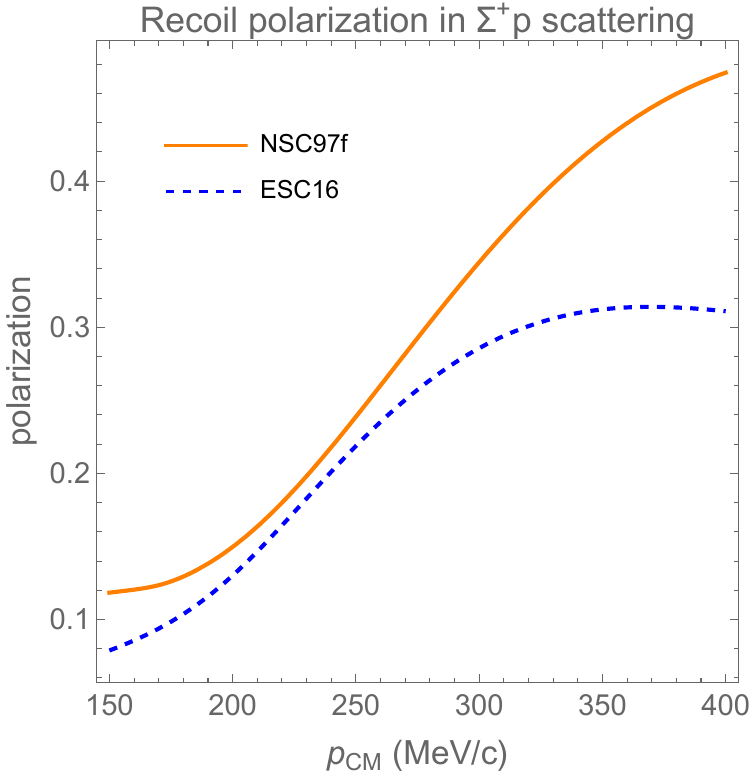}	\includegraphics[width=.23\textwidth]{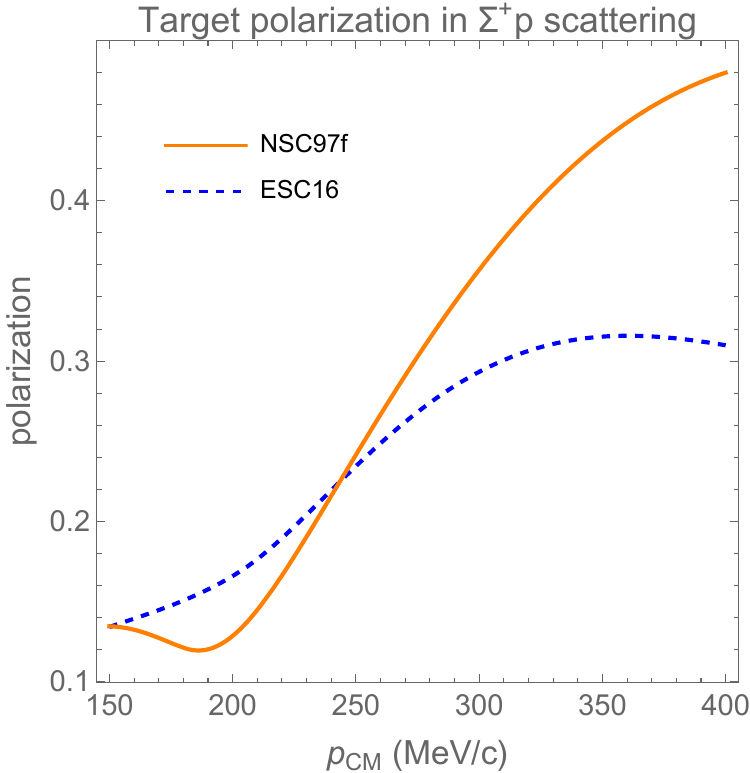}
	\caption{\em Predicted polarizations of the recoiling $\Sigma^+$ (recoil) and the recoiling $p$ (target) in $\Sigma^+ p$ scattering, assuming an unpolarized proton target and a 25\% polarized hyperon beam.
	}
	\label{fig:polfig}
\end{figure}

Given that, in the case of $\Sigma^+ p$,  different global fits  give rise to inconclusive outcome, we further propose ``quantum observables'' which could not only distinguish the varying global fits but also help determine the density matrix of the outgoing states \cite{workinprogress}. The observables are based on the formalism introduced in Ref.~\cite{Hoshizaki:1969qt}, which relates the density matrix of outgoing states to their polarizations. (For  quantum observables in top quark decays, see Refs.~\cite{Afik:2020onf,ATLAS:2023fsd}.) In Fig.~\ref{fig:polfig} we plot the predicted polarizations, as a function of $p_{CM}$, of  the recoiling hyperon (recoil polarization) and the recoiling proton (target polarization) in $\Sigma^+ p$ scattering from NSC97 and ESC16 fits, assuming an unpolarized proton target and a 25\% polarized incoming hyperon beam. We include phase shifts up to d wave as well as the $^3S_1-^3D_1$ mixing. We see that, by measuring the recoil and target polarizations, it is possible to distinguish between the two global fits. At J-PARC the $\Sigma^+$ particles come from the process $\pi^+p\to \Sigma^+K^+$ and is partially polarized in the order of 25\%, depending on the incoming momenta \cite{jparc}. A more detailed study of such a scenario will be presented elsewhere \cite{workinprogress}.

\section{Conclusion}

Hyperon-nucleon interactions are important for resolving the ``hyperon puzzle,'' which pertains to the formation of neutron stars heavier than two solar-masses. Inspired by recent experimental efforts in direct measurements of $YN$ scattering, as well as by theoretical advances in understanding nuclear dynamics from the perspective of quantum entanglement, we studied   in this work the information-theoretic properties of two-body $YN$ scatterings, focusing on the question of whether the spin entanglement is suppressed during the scattering process. Using globals fits of scattering data, we find hints of entanglement suppression in the majority of $YN$ scattering channels. In the case of $\Sigma^+ p$ scattering, only the more dated NSC97 fit showed entanglement suppression, while the more recent fits do not demonstrate entanglement suppression in the channel.

We further proposed polarizations of the recoiling hyperon and the recoiling proton as ``quantum observables'' which could  determine the entanglement property of the S-matrix. Such measurements could not only provide   access to the density matrix of the outgoing states, but also help differentiate the conflicting global fits, which will have important implications for the hyperon puzzle as well. In addition, these  observables open up new venues to investigate and measure the quantum nature of nucleons and hyperons.

\section*{Acknowledgements}

I.L. is supported in part by the U.S. Department of Energy under grant DE-SC0023522. Work at Argonne is supported in part by the U.S. Department of Energy under contract DE-AC02-06CH11357.  Discussions with Mikhail Bashkanov, Silas Beane,  Takuya Nanamura, Koji Miwa, Jen-Chieh Peng, Rik Yoshida and Nick Zachariou  are gratefully acknowledged.

\bibliography{YN_short_arxiv}

\end{document}